\documentclass[%
 reprint,
superscriptaddress,
 amsmath,amssymb,
 aps,
]{revtex4-1}

\usepackage{graphicx}
\usepackage{dcolumn}
\usepackage{bm}
\usepackage{xcolor}
\usepackage{mathrsfs}
\usepackage{soul}

\begin{document}

\title{Gain assisted controllable fast light generation in cavity magnomechanics}
\author{Sanket Das}
\affiliation{Department of Physics, Indian Institute of Technology Guwahati, Guwahati-781039, Assam, India.}
\author{Subhadeep Chakraborty}
\affiliation{Centre for Quantum Engineering Research and Education, TCG Centres for Research and Education in Science and Technology, Sector V, Salt Lake, Kolkata 70091, India}
\author{Tarak N. Dey}
\email{tarak.dey@iitg.ac.in}
\affiliation{Department of Physics, Indian Institute of Technology Guwahati, Guwahati-781039, Assam, India.}

\date{\today}

\begin{abstract}
We study the controllable output field generation from a cavity magnomechanical resonator system that consists of two coupled microwave resonators. The first cavity interacts with a ferromagnetic yttrium iron garnet (YIG) sphere providing the magnon-photon coupling. Under passive cavities configuration, the system displays high absorption, prohibiting output transmission even though the dispersive response is anamolous. We replace the second passive cavity with an active one to overcome high absorption, producing an effective gain in the system. We show that the deformation of the YIG sphere retains the anomalous dispersion. Further, tuning the exchange interaction strength between the two resonators leads to the system's effective gain and dispersive response. As a result, the advancement associated with the amplification of the probe pulse can be controlled in the close vicinity of the magnomechanical resonance. Furthermore, we find the existence of an upper bound for the intensity amplification and the advancement of the probe pulse that comes from the stability condition. These findings may find potential applications for controlling light propagation in cavity magnomechanics. \\

\end{abstract}

\maketitle

\section{\label{sec:level1}Introduction}
Cavity magnonics \cite{rameshti2022cavity, schlawin2022cavity}, has become an actively pursued field of research due to its potential application in quantum information processing \cite{Lachance-Quirion_2019,PhysRevB.93.174427}. The key constituent to such systems is a ferrimagnetic insulator with high spin density and low damping rate. It also supports quantized magnetization modes, namely, the magnons \cite{doi:10.1126/sciadv.1501286,PhysRevLett.113.083603}.	 	
With strongly coupled magnon-photon modes, cavity magnonics is an excellent platform for studying all the strong-coupling cavity QED effects \cite{PhysRevLett.113.156401}. Besides originating from the shape deformation of the YIG, the magnon can also couple to a vibrational or phonon mode \cite{doi:10.1126/sciadv.1501286}. This combined setup of magnon-photon-phonon modes, namely the cavity magnomechanics, has already demonstrated magnomechanically induced transparency \cite{doi:10.1126/sciadv.1501286}, magnon-induced dynamical backaction \cite{PhysRevX.11.031053}, magnon-photon-phonon entanglement \cite{PhysRevLett.121.203601,Li_2021}, squeezed state generation \cite{PhysRevA.99.021801},  magnomechanical storage and retreival of a quantum state \cite{Sarma_2021}.

Recently, $\mathcal{PT}$-symmetry drew extensive attention to elucidate the dynamics of a coupled system characterized by gain and loss \cite{PhysRevLett.80.5243,PhysRevLett.89.270401}.
Here, $\mathcal{P}$ stands for the parity operation, that results in an interchange between the two constituent modes of the system. 
The time reversal operator $\mathcal{T}$ takes $i$ to $-i$. $\mathcal{PT}$-symmetry demands the Hamiltonian is commutative with the joint $\mathcal{PT}$ operators {\it i.e.,} $[H, PT] = 0$.
This system possesses a spectrum of entirely real and imaginary eigenvalues that retain distinguishable characteristics \cite{El-Ganainy2018}. The point separating these two eigenvalues is the exceptional point (EP) \cite{Heiss_2012} where the two eigenvalues coalesce, and the system degenerates.  A natural testbed for $\mathcal{PT}$-symmetric Hamiltonian is optical as well as quantum optical systems \cite{PhysRevLett.100.103904,PhysRevLett.100.030402,PhysRevLett.103.093902} which already led to the demonstration of some of the exotic phenomena, like nonreciprocal light propagation \cite{Peng2014}, unidirectional invisibility \cite{Longhi_2011,Feng2013}, optical sensing and light stopping \cite{PhysRevLett.120.013901}. Very recently, a tremendous effort has been initiated to explore non-Hermitian physics in magnon assisted hybrid quantum systems. The second-order exceptional point is detected in a two-mode cavity-magnoic system, where the gain of the cavity mode is accomplished by using the idea of coherent perfect absorption \cite{Zhang2017}. The concept of Anti-$\mathcal{PT}$ symmetry has been realized experimentally \cite{PhysRevApplied.13.014053}, where the adiabatic elimination of the cavity field produces dissipative coupling between two magnon modes.
Beyond the unique spectral responses, these non-Hermitian systems can manipulate the output microwave field transmission \cite{Wang:23,PhysRevApplied.12.034001}. The underlying mechanism behind such an application is magnetically induced transparency\cite{PhysRevA.102.033721,doi:10.1126/sciadv.1501286}, where the strong magnon-photon coupling produces a narrow spectral hole inside the probe absorption spectrum. Further studies in this direction establish the importance of the weak magnon-phonon coupling to create double transmission windows separated by an absorption peak. Moreover, manipulating the absorption spectrum is also possible by varying the amplitude and phase of the applied magnetic field \cite{doi:10.1063/5.0028395}.

It is well established over the past decade that optomechanically induced transparency (OMIT)  \cite{PhysRevA.81.041803,Safavi-Naeini2011,doi:10.1126/science.1195596} is an essential tool for investigating slow light \cite{Hau1999} and light storage \cite{PhysRevLett.107.133601,PhysRevA.87.023812} in cavity. In addition, incorporating $\mathcal{PT}$-symmetry in optomechanical systems, provides a better controllability of light transmission \cite{Jing2015,Li2016} and produces subluminal to superluminal light conversion. Nonetheless, their proposals may find experimental challenges as the gain of the auxiliary cavity can lead the whole system to instability \cite{doi:10.1073/pnas.2019348118}. An eminent advantage of the magnomechanical system over the optomechanical system is that it offers strong hybridization between the magnon-photon mode. The magnomechanical systems offer better tunability as an external magnetic field can vary the magnon frequency. Exploiting these advantages, a $\mathcal{PT}$-symmetry-like magnomechanical system can be constructed by resonantly driving the YIG sphere to an active magnon mode \cite{PhysRevA.99.043803}. The controllable sideband generation with tunable group delay can be feasible by changing the power of the control field.

This paper investigates a controllable advancement and transmission of the microwave field from a coupled cavity magnomechanical system. Optical coupling between a passive cavity resonator containing YIG sphere and a gain-assisted auxiliary cavity can form a coupled cavity resonator. An external drive has been used to deform the YIG sphere's shape, resulting in the magnon-phonon interaction in the passive cavity. We show how the gain of the auxiliary cavity helps to overcome absorptive behaviour in our hybrid system. As a result, the output microwave field amplifies at the resonance condition. Moreover, the weak magnon-phonon interaction exhibits anomalous dispersion accompanied by a gain spectrum, demonstrating superluminal light. We also examine how the slope of the dispersion curve can be controlled by tuning the photon hopping interaction strength between the two cavities. 

The paper is organized as follows. In Section \ref{sec:theoretical model}, a theorical model for the coumpound cavity magnomechanical system with $\mathcal{PT}$-symmetric resonator is described. The Heisenserg equations of motion to govern the expectation values of operators of every system are derived in this Section. In Section \ref{sec:stability analysis}, we analyse the stability criteria of the model system and examine the effect of the auxiliary cavity gain on the absorptive and dispersive response of the system in Section \ref{sec:abs_disp}. Section \ref{sec:transmission} discusses the output probe field transmission. Further, the group
 velocity of the optical probe pulse has been studied analytically
 and verified numerically in Section \ref{sec:group delay}. Finally, we draw our conclusions in Section \ref{sec:conclusion}.
 
\section{\label{sec:theoretical model}Theoretical Model}
Recently, there has been a growing interest in realizing a gain in different components of cavity magnonics systems \cite{Zhang2017,PhysRevA.99.043803}. In this work, we investigate the effect of medium gain on the probe response and its transmission. The system under consideration is a hybrid cavity magnomechanical system that consists of two coupled microwave cavity resonators.
One of the resonators is passive and contains a YIG sphere inside it. We refer to this resonator as a cavity magnomechanical (CMM) resonator. Applying a uniform bias magnetic field to the YIG sphere excites the magnon mode. The magnon mode, in turn, couples with the cavity field by the magnetic-dipole interaction.
\begin{figure}[h!]%
	\centering
	{\includegraphics[scale=0.40]{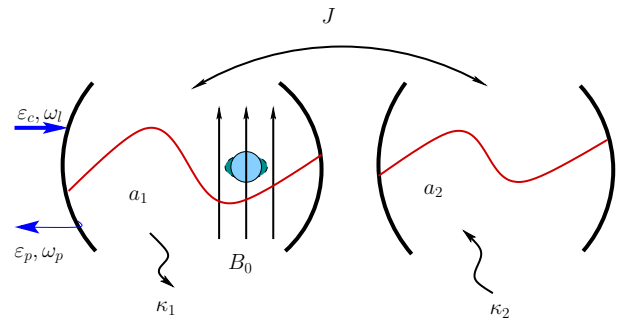} }
	\caption{\label{fig:model1}{The schematic diagram of a hybrid cavity magnomechanical system. The system consists of two coupled microwave cavities. One of them is passive, and another one is active. The passive cavity contains a ferromagnetic YIG sphere inside it. The applied bias magnetic field produces the magnetostrictive interaction between magnon and phonon. The coupling rates between the magnon-photon and magnon-phonon are $g_{ma}$ and $g_{mb}$, respectively. Strong control field of frequency $\omega_l$ and a weak probe field of frequency $\omega_p$ are applied to the passive cavity.}}
\end{figure}
 Nonetheless, the external bias magnetic field results in shape deformation of the YIG sphere, leading to the magnon-phonon interaction. The second resonator (degenerate with the first one) is coupled to the first resonator via optical tunnelling at a rate $J$. Two input fields drive the first resonator. The amplitude of the control, $\varepsilon_l$, and probe fields, $\varepsilon_p$, are given by $\varepsilon_i=\sqrt{P_i/\hslash\omega_i},(i\in l,p)$ with $P_i~\textrm{and}~\omega_i$ being the power and frequency of the respective input fields. The Hamiltonian of the combined system can be written as
\begin{align}
\label{eq:Hamiltonian}
 &H =\hslash \omega_c a_1^\dagger a_1 +\hslash \omega_c a_2^\dagger a_2+\hslash\omega_m m^\dagger m + \hslash\omega_b b^\dagger b\nonumber\\
 &+ \hslash J (a_1^\dagger a_2 + a_2^\dagger a_1)+\hslash g_{ma}(a_1^\dagger m + a_1 m^\dagger)\nonumber\\
 &+\hslash g_{mb}m^\dagger m(b^\dagger + b)+i\hslash\sqrt{2\eta_a\kappa_{1}}\varepsilon_l(a_1^\dagger e^{-i\omega_lt}-a_1e^{i\omega_lt})\nonumber\\
 &+i\hslash\sqrt{2\eta_a\kappa_{1}}\varepsilon_p(a_1^\dagger e^{-i\omega_p t}-a_1 e^{i\omega_p t}), 
\end{align}
where the first four terms of the Hamiltonian describe the free energy associated with each system's constituents.
The constituents of our model are characterized by their respective resonance frequencies: $\omega_c$ for the cavity mode, $\omega_m$ for the magnon mode, $\omega_b$ for the phonon mode. The annihilation operators for the cavity, magnon and phonon modes are represented by $a_i$, ($i=1,2$), $m$ and $b$, respectively. The fifth term signifies the photon exchange interaction between the two cavities with strength, $J$. The sixth term of the Hamiltonian corresponds to the interaction between the magnon and photon modes, characterized by a coupling rate $g_{ma}$. The interaction between the magnon and phonon modes is described by the seventh term of the Hamiltonian and the coupling rate between magnon and phonon mode is $g_{mb}$. Finally, the last two terms arise due to the interaction between the cavity field and two input fields. The cavity, magnon and phonon decay rates are characterized by $\kappa_1,\kappa_m~\textrm{and}~\kappa_b$, respectively. The coupling between the CMM resonator and the output port is given by $\eta_a=\kappa_{c_1}/2\kappa_1$, where $\kappa_{c_1}$ is the cavity external decay rate. In particular, we will consider the CMM resonator to be working in the critical-coupling regime where $\eta_a$ is $1/2$. At this point, it is convenient to move to a frame rotating at $\omega_l$. Following the transformation $H_{rot}=RHR^\dagger+ i\hslash\left({\partial R}/{\partial t}\right)R^\dagger$ with $R=e^{i\omega_l (a_1^\dagger a_1 + a_2^\dagger a_2 + m^\dagger m) t}$, the Hamiltonian in Eq. \eqref{eq:Hamiltonian} can be rewritten as
\begin{align}
\label{eq:rot_Hamiltonian}
    H_{rot}&=\hbar\Delta_a(a_1^\dagger a_1 + a_2^\dagger a_2) + \hbar\Delta_m m^\dagger m+\hbar\omega_b b^\dagger b \nonumber\\
    &+\hbar J (a_1^\dagger a_2 + a_2^\dagger a_1)+\hbar g_{ma}(a_1^\dagger m + a_1 m^\dagger)\nonumber\\
    &+\hslash g_{mb}m^\dagger m(b^\dagger + b)+i\hbar\sqrt{2\eta_a\kappa_{1}}\varepsilon_l(a_1^\dagger-a_1)\nonumber\\
    &+ i\hbar\sqrt{2\eta_a\kappa_{1}}\varepsilon_p(a_1^\dagger e^{-i\delta t}-h.c),
\end{align}
where $\Delta_a=\omega_c-\omega_l$ ($\Delta_m=\omega_m-\omega_l$) and $\delta = \omega_p-\omega_l$ are, respectively, the cavity (magnon) and probe detuning. The mean response of the system can be obtained by the Heisenberg - Langevin equation as $\langle\mathcal{\dot O}\rangle=i/\hbar\langle[H_{rot},\mathcal{O}]\rangle+\langle N \rangle$. Further, we consider the quantum fluctuations $(N)$ as white noise. Then starting form Eq. \ref{eq:rot_Hamiltonian}, the equations of motion of the system can be expressed as 
\begin{align}
\label{eq:langevin}
   \langle\dot{a_1}\rangle&=(-i\Delta_a-\kappa_1)\langle a_1\rangle-i g_{ma}\langle m\rangle -iJ\langle a_2\rangle \nonumber\\
   &+\sqrt{2\eta_a\kappa_1}\varepsilon_l+\sqrt{2\eta_a\kappa_1}\varepsilon_p e^{-i\delta t},\nonumber\\
   \langle\dot{m}\rangle&=(-i\Delta_m-\kappa_m)\langle m\rangle-ig_{ma}\langle a_1\rangle\nonumber\\
   &-ig_{mb}\langle m\rangle(\langle b^\dagger\rangle + \langle b\rangle),\nonumber\\
   \langle\dot{b}\rangle&=(-i\omega_b-\kappa_b)\langle b\rangle-ig_{mb}\langle m^\dagger\rangle \langle m\rangle,\nonumber\\
   \langle\dot{a_2}\rangle&=(-i\Delta_a+\kappa_2)\langle a_2\rangle-iJ\langle a_1\rangle,
\end{align}
where $\kappa_{2}$ and $\kappa_b$ respectively denote the gain of the second resonator and phonon damping rates. We note that $\kappa_{2}>0$ corresponds to a coupled passive-active CMM resonators system and $\kappa_{2}<0$ describes a passive-passive coupled CMM resonators system. Assuming the control field amplitude $\varepsilon_l$ to be larger than the probe field $\varepsilon_p$, each operator expectation values $\langle\mathcal{O}(t)\rangle$ can be decomposed into its steady-state values $\mathcal{O}_s$ and a small fluctuating term $\delta\mathcal{O}(t)$. The steady-state values of each operator are
\begin{subequations}
\begin{align}
\label{eq:steady_state}
a_{1s}&=\frac{(-i\Delta_a+\kappa_2)(-ig_{ma}m_s+\sqrt{2\eta_a\kappa_1}\varepsilon_l)}{(i\Delta_a+\kappa_1)(-i\Delta_a+\kappa_2)-J^2},\\
m_s&=\frac{-ig_{ma}a_{1s}}{i\Delta_m'+\kappa_m}, \\
b_s&=\frac{-ig_{mb}|m_s|^2}{i\omega_b+\kappa_b}, \\
a_{2s}&=\frac{iJa_{1s}}{(-i\Delta_a+\kappa_2)}.
\end{align}
\end{subequations}
While the fluctuating parts of Eq. \ref{eq:langevin} can be expressed as 
\begin{align}
\label{eq:fluctuation}
\delta \dot a_1&=-\left(i\Delta_a+\kappa_1\right)\delta a_1-iJ\delta a_2-i g_{ma}\delta m\nonumber\\
&+\sqrt{2\eta_a\kappa_1}\varepsilon_p e^{-i\delta t},\nonumber\\
\delta\dot{m}&=-(i\Delta'_m+\kappa_m)\delta m-ig_{ma}\delta a_1 -iG\delta b-iG\delta b^\dagger,\nonumber\\
\delta \dot b&=-\left(i\omega_b+\kappa_b\right)\delta b-iG\delta m^\dagger-iG^*\delta m,\nonumber\\
\delta \dot a_2&=-\left(i\Delta_a-\kappa_2\right)\delta a_2-iJ\delta a_1,
\end{align}
where $\Delta_m'=\Delta_m+g_{mb}(b_s+b_s^*)$ is the effective magnon detuning and $G=g_{mb}m_s$ is the enhanced magnon-phonon coupling strength. For simplicity, we express these fluctuation equations as
 \begin{align}
 \label{eq:shrodinger_eq}
   i\frac{d}{dt}|\psi\rangle&=H_{eff}|\psi\rangle + F,
 \end{align}
 where the fluctuation vector $|\psi\rangle=(\delta a_1,\delta a_1^\dagger,\delta a_2,\delta a_2^\dagger,\delta b,\delta b^\dagger,\delta m,\delta m^\dagger)^T$, input field $F=(\sqrt{2\eta_a\kappa_1}\varepsilon_p e^{-i\delta t},\sqrt{2\eta_a\kappa_1}\varepsilon_p e^{i\delta t},0,0,0,0,0,0)^T$.
Next, we adopt the following ansatz to solve Eq. \ref{eq:fluctuation}:
\begin{align}
\label{eq:ansatz}
\delta a_1(t)&=A_{1+}e^{-i\delta t} + A_{1-}e^{i\delta t},\nonumber\\
\delta m(t)&=M_{+}e^{-i\delta t} + M_{-}e^{i\delta t}\nonumber\\
\delta b(t)&=B_{+}e^{-i\delta t} + B_{-}e^{i\delta t},\nonumber\\
\delta a_2(t)&=A_{2+}e^{-i\delta t} + A_{2-}e^{i\delta t}.
\end{align}
Here $A_{i+}~\textrm{and}~A_{i-}$ correspond to the $i^{th}$ cavity generated probe field amplitude and the four-wave mixing field amplitude, respectively.
By considering $h_1=-i\Delta_a+i\delta-\kappa_1,~h_2=-i\Delta_a-i\delta-\kappa_1,~h_3=-i\Delta_a+i\delta+\kappa_2,~h_4=-i\Delta_a-i\delta+\kappa_2,~h_5=-i\omega_b+i\delta-\kappa_b,~h_6=-i\omega_b-i\delta-\kappa_b,~h_7=-i\Delta_m'+i\delta-\kappa_m,~h_8=-i\Delta_m'-i\delta-\kappa_m$,
we obtain $A_{1+}$ which corresponds to the output probe field amplitude from the CMM resonator as 
\begin{align}
A_{1+}(\delta)=\frac{C(\delta)}{D(\delta)},
\end{align}
where
\begin{align}
 \label{eq:coefficient}
 &C(\delta)=-\sqrt{2\eta_a\kappa_a}\varepsilon_p h_3(h_5 h_7 h_6^*(J^2h_8^*+h_4^*(g_{ma}^2+h_2^*h_8^*))\nonumber\\
 &+|G|^2(h_5-h_6^*)(J^2(h_7-h_8^*)-h_4^*(g_{ma^2}+h_2^*(h_8^*-h_7^*)))),\nonumber \\
&D(\delta)=h_5h_6^*(g_{ma}^2h_3+h_7(h_1h_3+J^2))\nonumber\\
&(J^2h_8^*+h_4^*(g_{ma}^2+h_2^*h_8^*))
+|G|^2(h_5-h_6^*)\nonumber\\
&(J^2(g_{ma}^2h_3+(h_1h_3+J^2)(h_7-h_8^*))
-h_4^*\nonumber\\
&((h_1h_3+J^2)
(g_{ma}^2-h_2^*(h_7-h_8^*))-h_2^*h_3g_{ma}^2)).
\end{align}
The output field from the CMM resonator is obtained by the cavity input-output relation
\begin{align}
\label{eq:input-output}
\varepsilon_{out}&=\sqrt{2\eta_a\kappa_1}\langle a_1\rangle-\varepsilon_l-\varepsilon_p e^{-i\delta t}.
\end{align}
By substituting Eq. \ref{eq:ansatz} into Eq. \ref{eq:input-output}, we obtain the normalized output probe field intensity from the CMM resonator as
\begin{align}
\label{eq:output_power}
T=|t_p|^2&=\left|\frac{\sqrt{2\eta_a\kappa_1}A_{1+}}{\varepsilon_p}-1\right|^2.
\end{align}
In order to numerically simulate the transmitted output probe field spectrum, we use the following experimentally realizable set of parameter values \cite{PhysRevB.105.214418,doi:10.1126/sciadv.1501286}. The degenerate microwave cavities of frequency $\omega_c/2\pi=7.86$ GHz. The decay rate of the first cavity is $\kappa_1/\pi=3.35$ MHz. The spin density $\rho=4.22\times10^{27}$ m$^{-3}$ and the diameter of the YIG sphere $D=25$ $\mu$m. It results in $3\times10^{16}$ number of spins ($N_m$) present in the YIG sphere. The phonon mode has frequency $\omega_b/2\pi=11.42$ MHz with decay rate $\kappa_b/\pi=300$ Hz, and the magnon-phonon coupling strength $g_{mb}/2\pi$ is $1$ Hz. The Kittel mode frequency of the YIG sphere is $\omega_m=\gamma_e B_{0,i}$, with gyromagnetic ratio, $\gamma_e/2\pi=28$ GHz/T and $B_{0,i}$ is the input bias magnetic field amplitude. The magnon decay rate is $\kappa_m=3.52$ MHz. Magnon-photon coupling strength $g_{ma}=\gamma_e B_{vac}\sqrt{5N_m}/2$ can be controlled by changing the vacuum magnetic field amplitude as $B_{vac}=\sqrt{2\pi\hbar\omega_c/V}$.\\
\section{\label{sec:Results}Results}
\subsection{\label{sec:stability analysis}Stability Analysis}
\begin{figure}[t!]%
 	\centering
 	{\includegraphics[scale=0.33]{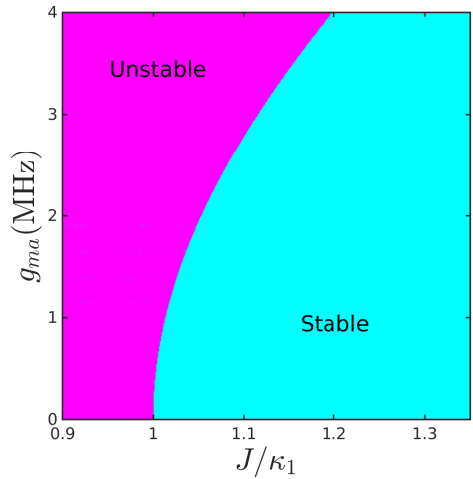}%
 		\caption{\label{fig:stability1}The stable and unstable regions are determined as a function of normalized evanescent coupling strength ($J/\kappa_1$) and the cavity-magnon coupling strength ($g_{ma}$) when the loss of the CMM-resonator is perfectly balanced by the gain of the auxiliary cavity $(\kappa_1=\kappa_2)$. We consider the	control field intensity to be $10$ mW. The other parameters are $\omega_c=2\pi\times 7.86$ MHz, $\omega_b=2\pi\times11.42$ MHz, $\Delta_a=\Delta_m'=\omega_b=2\pi\times11.42$ MHz, $\kappa_1=\kappa_2=\pi\times 3.35$ MHz, $\kappa_m=\pi\times1.12$ MHz, $\kappa_b=\pi\times300$ Hz, and $g_{mb}=2\pi$ Hz. }}
 \end{figure}

Initially we consider the two coupled cavities which are operating under a balanced gain-loss condition. The Hamiltonian describing such coupled resonator system ($g_{ma}=g_{mb}=0$) can be written as 
\begin{align}
	\label{eq:pt_hamiltonian}
H_{cav}&=\hbar(\Delta_a - i\kappa_1) \delta a_1^\dagger \delta a_1 + \hbar(\Delta_a + i\kappa_1) \delta a_2^\dagger \delta a_2\nonumber\\
&+ \hbar J (\delta a_1^\dagger \delta a_2 + \delta a_2^\dagger \delta a_1 ).
\end{align}
 The eigenvalues of $H_{cav}$ are $\lambda_{\pm}=\Delta_a\pm\sqrt{J^2-\kappa_1^2}$. Note that the above Hamiltonian remains invariant under the simultaneous parity $\mathcal{P}:a_1\leftrightarrow a_2$ and time-reversal operation $\mathcal{T}: i\rightarrow-i$ operations, and, its eigenvalues are entirely real and complex for $J>\kappa_1$ and $J<\kappa_2$. The point $J=\kappa_1$, which marks this transition from $\mathcal{PT}$ symmetric to the $\mathcal{PT}$ breaking phase, is known as the exceptional point (EP). One must understand the competitive behaviour between the inter-cavity field coupling and the loss/gain rates to get insight into this transition. For $J>\kappa_1$ the intracavity field amplitudes can be coherently exchanged and thus give rise to a coherent oscillation between the field amplitudes. However, for $J<\kappa_1$ the intracavity field can not be transferred to the other one, resulting in a strong field localization or in other words exponential growth. A quick look at Eq. 4(a) also suggests such gain-induced dynamic instability in $a_1$ at $J=\kappa_1$ for $\Delta_a =0$. This situation becomes more complicated in the presence of magnon-photon coupling. Now, the combined system ($g_{ma}, g_{mb} \neq 0$) ceases to become $\mathcal{PT}$ symmetric. However, the effect of an additional gain cavity ($\kappa_2>0$) can be understood by analyzing the stability diagram of the whole system. In the following, we derive the stability condition by invoking the Routh-Hurwitz criterion which requires all the eigenvalues of $H_{eff}$ have negative real parts. The magenta region of Fig. \ref{fig:stability1} suggests that when $g_{ma}$ is small the instability threshold remains close to the $J=\kappa_1$ (the conventional EP for a binary $\mathcal{PT}$ symmetric system). However, with increasing $g_{ma}$ the system reaches instability at a larger exchange interaction $J$. Such a restriction over the choice of the photon exchange rate parameter $J$ will be followed throughout this paper.
\\
\subsection{\label{sec:abs_disp}\textbf{Absorption and dispersion spectrum}}
\begin{figure}[h!]%
	\centering
	{\includegraphics[scale=0.32]{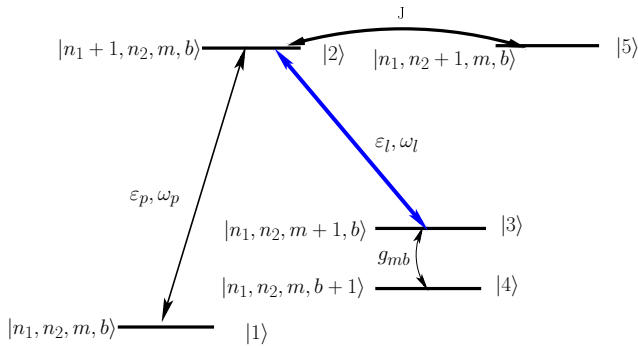} }
	\caption{\label{fig:level-diagram}{Level diagram of the model system. $|n_i\rangle$,$|m\rangle~\textrm{and}~|b\rangle$ represents the photon number state of $i^{th}$ cavity, magnon mode and phonon mode, respectively. The application of strong control field to the CMM resonator couples $|n_1+1,n_2,m,b\rangle$} and $|n_1,n_2,m+1,b\rangle$, whereas, the presence of a weak probe field increases the photon number of CMM resonator by unity. $g_{mb}$ couples $|n_1,n_2,m+1,b\rangle~\textrm{and}~|n_1,n_2,m,b+1\rangle$. The hopping interaction between the two cavities directly couples $|n_1+1,n_2,m,b\rangle\leftrightarrow|n_1,n_2+1,m,b\rangle$. }
\end{figure}

\begin{figure}[h!]
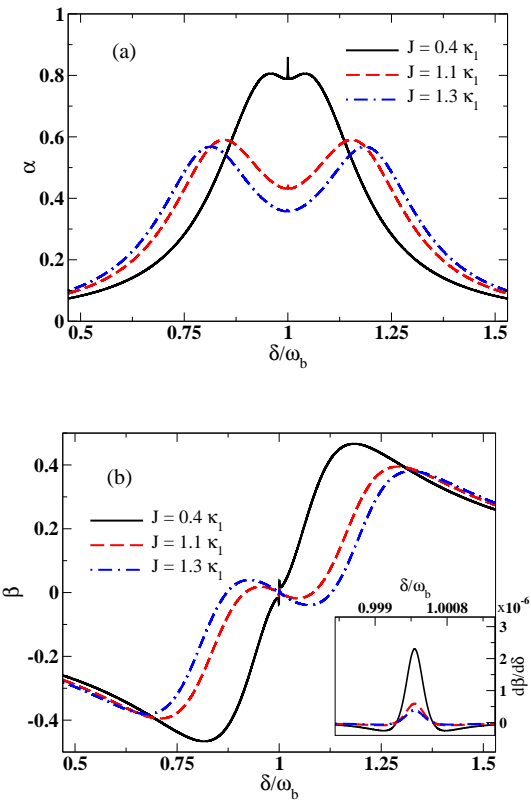
%
	\centering
	{\includegraphics[scale=0.23]{fig4a.eps} }
	
	\vspace{0.80cm}
	
	{\includegraphics[scale=0.23]{fig4b.eps} }%
	\vspace{0.1cm}
	\caption{\label{fig:absorption}(a) Absorption and (b) dispersion spectrum of the model system. The slope of the dispersion curve is shown in the inset. Here we consider both the microwave cavities are passive, with identical decay rates $(\kappa_1=-\kappa_2)$. The magnon-photon couplong strength, $g_{ma}$ is taken as $2$ MHz. All the other parameters are mentioned earlier.}%
\end{figure}

The magnomechanical system under consideration corresponds to the level diagram of Fig. \ref{fig:level-diagram}. Application of a probe field excites the passive cavity mode and allows the transition between $|1\rangle~\textrm{and}~|2\rangle$. The exchange interaction, $J$, couples two degenerate excited states $|2\rangle$ and $|5\rangle$. The presence of the strong control field distributes the population between the two states $|2\rangle$ and $|3\rangle$. The magnon-phonon coupling, $g_{mb}$, couples both the metastable ground states $|3\rangle$ and $|4\rangle$. Here, we consider both the microwave cavities to be passive ($\kappa_2<0$) under a weak magnon-photon coupling strength, $g_{ma}$. In this situation, the magnon-photon hybridization is insignificant. The absorptive and dispersive response can be quantified by the real and imaginary components of $(t_p+1)$ that will be presented as $\alpha$ and $\beta$, respectively. In Fig. \ref{fig:absorption}(a), we present the absorptive response of the system as a function of normalized probe detuning. The black-solid curve depicts a broad absorption spectrum of the probe field when the exchange interaction is much weaker than the cavity decay rate.
One can explain it by considering the level diagram of Fig. \ref{fig:level-diagram}, where the initial population stays in the ground state $|1\rangle$. Applying a probe field transfers the population from the ground state to the excited state, $|2\rangle$. In addition, the weak magnon-photon coupling (with respect to $\kappa_1$) restricts a significant transition from $|2\rangle$ to $|3\rangle$. As a result, it allows the transfer of a fraction of the excited state’s population by invoking the exchange interaction  $J$.
 The increase in the exchange interaction strength causes a gradual decrease in $|2\rangle$'s population. It reduces the absorption coefficient around the resonance condition except for $\delta=\omega_b$. This phenomenon is shown by the red-dashed and the blue dotted-dashed curve of Fig. \ref{fig:absorption}(a). We observe a narrow absorption peak inside the broad absorption peak for $J=0.4$ $\kappa_1$. The sharp absorption peak, exactly at $\delta=\omega_b$, occurs due to the magnomechanical resonance. Further increasing the exchange interaction virtually cuts off the population distribution from $|2\rangle$ to $|3\rangle$. As a result, the effect of magnon-phonon resonance also decreases, and the absorption peak at $\delta=\omega_b$ eventually diminishes. In Fig. \ref{fig:absorption}(b), we present the dispersion spectrum as a function of normalized detuning $\delta/\omega_b$. For the time being, we neglect the effect of magnomechanical coupling and observe the occurrence of anomalous dispersion around $\delta=\omega_b$ for $J = 0.4$ $\kappa_1$. Further, increasing the exchange interaction strength more significant than the cavity decay rate can alter the dispersive response from anomalous to normal, as shown by the red-dashed and blue-dot-dashed curves.
In the inset of Fig. \ref{fig:absorption}(b), we plot the slope of the temporal dispersion $d\beta/d\delta$ at the extreme vicinity of the magnon-phonon resonance condition. The positive values of the slope of the temporal dispersion signify anomalous dispersion due to the magnomechanical coupling. However, the steepness of the dispersion curve can be reduced by increasing the exchange interaction strength, as shown by the red-dashed and blue-dot-dashed curves. Note that this dispersion curve is accompanied by absorption. Output transmission of the probe field is prohibited in the presence of huge absorption. Therefore, reducing absorption or introducing the gain to the system is mandatory for observing the group velocity phenomena.\\
\begin{figure}[h!]
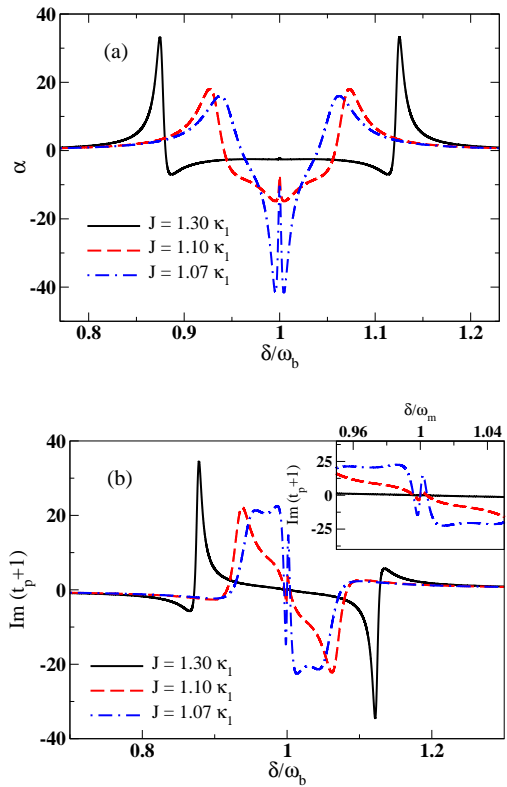
%
	\centering
	{\includegraphics[scale=0.23]{fig5a.eps} }
	
	\vspace{0.40cm}
	
	{\includegraphics[scale=0.23]{fig5b.eps} }%
	\vspace{0.1cm}
	\caption{\label{fig:a_p_abs_disp}(a) Absorption and (b) dispersion spectrum of the model system. The slope of the dispersion curve is shown in the inset. Here we consider the second cavity as a gain cavity, with  $\kappa_2=\kappa_1$. All the other parameters are the same as before.}%
\end{figure}
To achieve reasonable transmission at the output, we replace the auxiliary passive cavity with an active one where the second cavity's gain ($\kappa_2>0$) completely balances the first cavity's loss. In this scenario, the stability criterion for the hybrid system allows us to consider the exchange interaction strength $J$ to be greater than $1.053$ $\kappa_1$ for $g_{ma}=2$ MHz. We present the absorptive response of the model system in Fig. \ref{fig:a_p_abs_disp}(a). The black solid curve of Fig. \ref{fig:a_p_abs_disp}(a) illustrates the occurrence of a double absorption peak spectrally separated by a broad gain regime. The graphical nature is determined by the roots of $D(\delta)$, which are, in general, complex. The real parts of the roots determine the spectral peak position, and the imaginary parts correspond to their widths. To illustrate this, we consider $J=1.30$ $\kappa_1$ with all other parameters remaining the same as earlier. The real parts of the root of $D(\delta)$ present two distinct normal mode positions at $\delta/\omega_b$ values $0.88~\textrm{and}~1.12$. The other two normal modes are spectrally located at the same position $\delta/\omega_b=1$. The interference between these two normal modes becomes significant while approaching the stability bound as depicted by the red dashed and blue dot-dashed curve of Fig. \ref{fig:a_p_abs_disp}(a). In turn, it reduces the overall gain of the composite system. Further, we investigate the effect of a gain-assisted auxiliary cavity on the medium's dispersive response in Fig. \ref{fig:a_p_abs_disp}(b). For $J=1.30$ $\kappa_1$, the two absorption peaks produce two distinct anomalous dispersion regions separated by a broad normal dispersive window. Weakening the exchange interaction strength reveals prominent normal dispersion around the resonance condition except for $\delta=\omega_b$, and the window shrinks. In the inset of Fig. \ref{fig:a_p_abs_disp}(b), we present the slope of the dispersive response due to the magnomechanical resonance. The black solid curve of Fig. \ref{fig:a_p_abs_disp}(b) suggests the occurrence of anomalous dispersion at the magnon-phonon resonance condition. Moreover, one can increase the steepness of the dispersion curve by simply approaching the instability threshold, as delineated by the red-dashed and blue-dotted-dashed curve of the inset of Fig. \ref{fig:a_p_abs_disp}(b). In the consecutive section, we will discuss how the change in the dispersion curve can produce controllable group velocity of the light pulses through the medium and investigate the role of the exchange interaction.
\subsection{\label{sec:transmission}\textbf{Output probe transmission}}
\begin{figure}[b!]
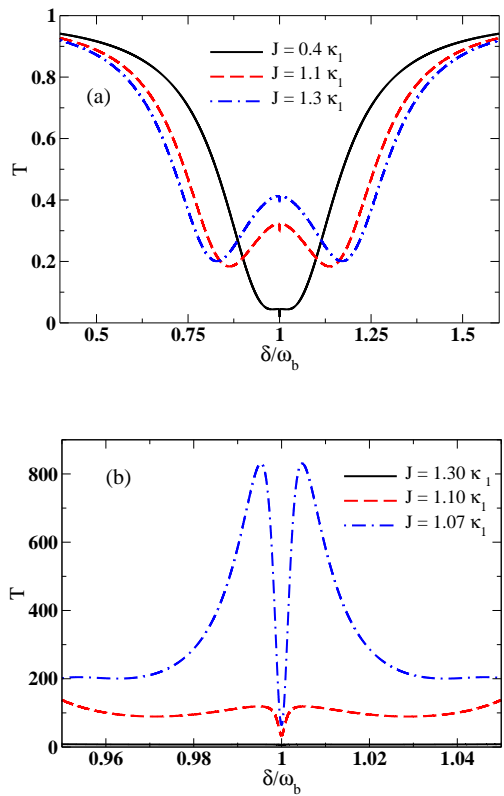
%
	\centering
	{\includegraphics[scale=0.23]{fig6a.eps} }
		
	\vspace{0.87cm}
		
	{\includegraphics[scale=0.23]{fig6b.eps} }%
	\vspace{0.1cm}
	\caption{\label{fig:transmission}{Exchange interaction $J$ dependent normalized output probe transmission 
	is plotted as a function of normalized detuning between the control and the probe field when (a) both the cavities are passive ones, and (b) one is active and another one is passive.}}%
	\end{figure}

The output probe intensity from the system depends on its absorptive response. Equation \ref{eq:output_power} dictate the transmission of the probe field and is presented in Fig. \ref{fig:transmission}. For Fig. \ref{fig:transmission}(a), we consider both the microwave cavities as passive ones with identical decay rates, {\it i.e.,} $\kappa_1=-\kappa_2$. The black solid curve shows a broad absorptive response for $J=0.40$ $\kappa_1$. Increasing the exchange interaction strength causes gradual enhancement in the output probe transmission, as delineated by the red-dashed and blue dot-dashed curve of Fig. \ref{fig:transmission}(a), and the absorption window splits into two parts. A precise observation confirms the presence of extremely weak transmission dip exactly at $\delta=\omega_b$ for all the three exchange interaction strengths under consideration.
	 In Fig. \ref{fig:transmission}(b), we present the advantage of using a gain-assisted auxiliary cavity along with a CMM resonator to obtain a controllable amplification of the output probe field. We begin our discussion considering the photon hopping interaction, $J=1.30$ $\kappa_1$. The black solid curve of Fig. \ref{fig:transmission}(b) estimates the normalized probe transmission of $6.03$. Here the normalization is done with respect to the input probe field intensity. By decreasing the parameter $J$, we approach the unstable region and observe the occurrence of a double transmission peak separated by a sharp and narrow transmission dip. The amplitude of the double transmission peak demonstrates the probe pulse amplification by a factor of $830$, as presented by the blue dotted-dashed curve.  However, an explicit observation suggests the output probe field amplification by a factor of $67$ at the resonance condition $\delta=\omega_b$. The physics behind the probe field amplification can be well understood as: Introduction of gain to the second cavity compensates a portion of losses in the first cavity through $J$. This leads to an enhanced field amplitude in the first cavity. In the presence of moderate magnon-photon coupling it increases the effective magnon-photon coupling strength. Hence, we observe a higher transmission at the two sidebands but also find a large transmission dip at $\delta=\omega_b$.
\subsection{\label{sec:group delay}Group delay}
 \label{sec:Group delay}
Controllable group delay has gained much attention due to its potential application in quantum information processing and communication. The dispersive nature of the medium is the key to controlling the group delay of the light pulse under the assumption of low absorption or gain. The pulse with finite width in the time domain is produced by superposing several independent waves with different frequencies centered around a carrier frequency $(\omega_s)$. The difference in time between free space propagation and a medium propagation for the same length can create a group delay. The analytical expression for the group delay can be constructed by considering the envelope of the optical pulse as
\begin{equation}
 f(t_0)=\int_{-\infty}^{\infty}\tilde{f}(\omega)e^{-i\omega t_0}d\omega,\nonumber\\
\end{equation}
where $\tilde f(\omega)$ corresponds to the envelope function in the frequency domain. Accordingly, the reflected output probe pulse can be expressed as
\begin{align}
 \label{eq:definition}
 f^R(t_0)=&\int_{-\infty}^{\infty}t_p(\omega)\tilde{f}(\omega)e^{-i\omega t_0}d\omega,\\  
 =&e^{-i\omega_s t_0}\int_{-\infty}^{\infty}t_p(\omega_s+\delta)\tilde{f}(\omega_s+\delta)e^{-i\delta t_0}d\delta,\nonumber\\
 =&t_{p}(\omega_s)e^{-i\omega_s\tau_{g}}f(t_0-\tau_{g}).
\end{align}
This expression can be obtained by expanding $t_p(\omega_s+\delta)$ in the vicinity of $\omega_s$ by a Taylor series and keeping the terms upto first order in $\delta$. An expression for time-delay is obtained as \cite{Safavi-Naeini2011,Das:22}
\begin{equation}
  \label{eq:timedelay}
  \begin{aligned}
    \tau_g= {\textrm Re}\left[\frac{-i}{t_p(\omega_s)}\left(\frac{dt_p}{d\omega}\right)\bigg|_{\omega_s}\right],\\       
  \end{aligned}
\end{equation}
which can be further simplified as
\begin{equation}
  \label{eq:timedelay1}
  \begin{aligned}
    \tau_g= \frac{{\left(\alpha(\omega_s)-1\right)}{\frac{d\beta}{d\omega}}\bigg|_{\omega_s}-{\beta(\omega_s)}{\frac{d\alpha}{d\omega}}\bigg|_{\omega_s}}{|t_p(\omega_s)|^2}.      
  \end{aligned}
\end{equation}
From Eq. \ref{eq:timedelay1}, the slope of the medium's absorption and dispersion curves determine the probe pulse propagation delay or advancement. However, Fig. \ref{fig:a_p_abs_disp}(b) suggests that the value of $\beta$ is negligibly small near the magnomechanical resonance. Hence, the group delay depends on the first term of the numerator of Eq. \ref{eq:timedelay1}.
\begin{figure}[t!]%
    \centering
    {\includegraphics[scale=0.23]{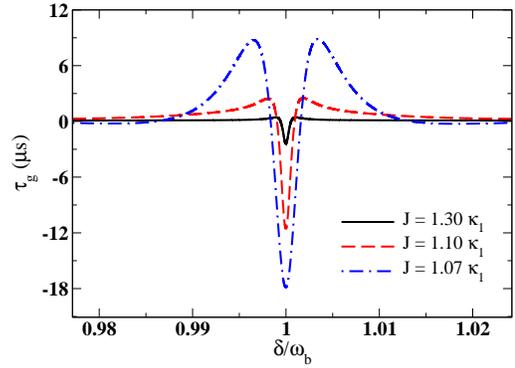} }%
    \caption{\label{fig:delay}Time delay of the probe pulse for different evanescent coupling strength $J$ have been plotted against the normalized probe detuning $\delta/\omega_b$, while the control power is $10$ mW. All other parameters are
taken as the same as in Fig. \ref{fig:a_p_abs_disp}.}%
\end{figure}
 In Fig. \ref{fig:delay}, we examine the effect of photon-photon exchange interaction on the probe pulse propagation delay when both cavities operate under balanced gain-loss condition. The system produces anomalous dispersion accompanied by a gain response. The black solid curve of Fig. \ref{fig:delay} depicts the probe pulse advancement of $2.4$ $\mu$s for the photon hopping interaction strength, $J=1.3$ $\kappa_1$. Moreover, one can enhance the effective gain and the slope of the anomalous dispersion curve by approaching the instability threshold. That, in turn, brings out the super luminosity of the output probe pulse, characterized by the advancement of $17.9$ $\mu$s as shown by the blue dotted-dashed curve of Fig. \ref{fig:delay}.
 
To verify the above results, we consider a Gaussian probe pulse with a finite width around the resonance condition, {\it i.e.,} $\delta=\omega_b$, and numerically integrate it by using Eq. \ref{eq:definition}. The shape of the input envelope is considered as,
\begin{equation}
 \label{eq:shape}
 \tilde{f}(\omega)=\frac{\varepsilon_p}{\sqrt{\pi\Gamma^2}}e^{-\frac{(\omega-\omega_s)^2}{\Gamma^2}},
\end{equation}
where $\Gamma$ is the spectral width of the optical pulse. We consider $\Gamma$ to be $7.17$ kHz, such that the Gaussian envelope is well-contained inside the gain-window around the resonance condition ($\delta=\omega_b$), as depicted in Fig. \ref{fig:a_p_abs_disp}(a).
 The dispersive, absorptive as well as gain response of the system can be demonstrated by examining the effect of the probe transmission coefficient ($t_p$) on the shape of the input envelope.
 The gain of the auxiliary cavity manipulates the probe transmission coefficient in such a way that it amplifies the intensity of the output probe pulse. The black solid curve depicts the output probe pulse amplification of $6.2$ for photon-hopping interaction strength, $J=1.30$ $\kappa_1$. A decrease in the $J$ value gradually enhances the effective gain in the system.
 \begin{figure}[h!]%
 	\centering
 	{\includegraphics[scale=0.23]{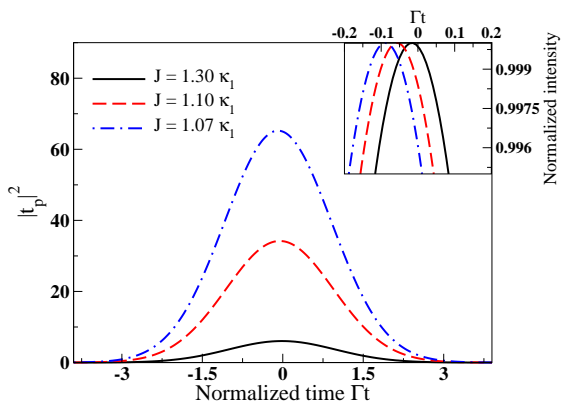} }
 	\caption{\label{fig:pulse-prop}{The relative intensity of the output probe pulse is plotted against the normalized time ($\Gamma$t) for different photon-photon exchange interaction strength when both cavities are operating under balanced gain-loss condition.}}
 \end{figure}
  It amplifies the output probe transmission as presented by the red dashed and blue dotted-dashed curves of Fig. \ref{fig:pulse-prop}. We observe that the output field amplification can reach to a factor of 65.3 while considering the exchange interaction strength to be $1.07$ $\kappa_1$.
  Further decreasing the exchange interaction will lead to dynamical instability in our model system. Interestingly, the temporal width of the probe pulse is almost unaltered during the propagation through the magnon-assisted double cavity system. This numerical result agrees with our analytical results for the output probe transmission, as shown in Fig. \ref{fig:transmission}(b). Moreover, the importance of the photon-photon exchange interaction on the probe pulse propagation advancement can be observed from the inset of Fig. \ref{fig:pulse-prop}. The peak separation between the input pulse $(t=0)$ and the output pulse for $J=1.30$ $\kappa_1$ estimates the probe pulse advancement of $2.34$ $\mu$s. The red dashed, and blue dashed-dotted curve of the inset estimates the probe pulse advancement of $8.75$ $\mu s$ and $13.30$ $\mu s$ for $J=1.10$ $\kappa_1~\textrm{and}~1.07$ $\kappa_1$, respectively.\\
 
\section{\label{sec:conclusion}Conclusion}
In conclusion, we have theoretically investigated the controllable output field transmission from a critically coupled cavity magnomechanical system. We drive the first cavity with a YIG sphere inside it, establishing the magnon-photon coupling. The photon exchange interaction connects the second microwave cavity with the first. An external magnetic field induces the deformation effect of the YIG sphere. In this study, the interaction between the magnon and photon modes lies under the weak coupling regime. The medium becomes highly absorbent when both cavities are passive, and the output probe transmission is prohibited. We introduce a gain to the auxiliary cavity to overcome this situation. It is noteworthy that the instability threshold must be close to the conventional exceptional point for a binary $\mathcal{PT}$-symmetric system. At the magnomechanical resonance, the auxiliary cavity produces an effective gain associated with anomalous dispersion.
Further, decreasing the photon exchange interaction strength causes gradual enhancement of the effective gain and the steepness of the dispersion spectrum. As a result, we observe a controllable superluminal microwave pulse propagation associated with amplification by a factor of 67. By studying the propagation dynamics of a Gaussian probe pulse of width $7.17$ kHz, we confirm that the numerical study is consistent with the analytical results. Our study may find potential applications in weak signal sensing and communication purposes in the newly emerging field of cavity magnomechanics.
\nocite{*}
\bibliography{reference}
\end{document}